\documentclass[aps,prl,nobibnotes,twocolumn,superscriptaddress,bibliography]{revtex4-1}
\usepackage{amsfonts}
\usepackage{mathrsfs}
\usepackage{amsmath}
\usepackage{color}
\usepackage{graphicx}
\usepackage{bm}
\usepackage{amssymb}
\usepackage{xspace}
\usepackage{epstopdf}
\usepackage{dcolumn}
\usepackage{longtable}
\usepackage{multirow}
\usepackage{float}
\usepackage{comment}
\makeatletter

\providecommand{\tabularnewline}{\\}

\usepackage[colorlinks=true, letterpaper=true, pdfstartview=FitV, linkcolor=blue, citecolor=blue, urlcolor=blue]{hyperref}

\makeatother
\begin{document}
\title{Hidden Real Topology and Unusual Magnetoelectric Responses in Monolayer Antiferromagnetic Cr$_2$Se$_2$O}

\author{Jialin Gong}
\address{School of Physical Science and Technology, Southwest University, Chongqing 400715, China}

\author{Yang Wang}
\affiliation{Key Lab of advanced optoelectronic quantum architecture and measurement (MOE), Beijing Key Lab of Nanophotonics $\&$ Ultrafine Optoelectronic Systems, and School of Physics, Beijing Institute of Technology, Beijing 100081, China}

\author{Yilin Han}
\affiliation{Key Lab of advanced optoelectronic quantum architecture and measurement (MOE), Beijing Key Lab of Nanophotonics $\&$ Ultrafine Optoelectronic Systems, and School of Physics, Beijing Institute of Technology, Beijing 100081, China}

\author{Zhenxiang Cheng}
\address{Institute for Superconducting and Electronic Materials (ISEM), University of Wollongong, Wollongong 2500, Australia}

\author{Xiaotian Wang}
\address{School of Physical Science and Technology, Southwest University, Chongqing 400715, China}
\address{Institute for Superconducting and Electronic Materials (ISEM), University of Wollongong, Wollongong 2500, Australia}

\author{Zhi-Ming Yu}
\affiliation{Key Lab of advanced optoelectronic quantum architecture and measurement (MOE), Beijing Key Lab of Nanophotonics $\&$ Ultrafine Optoelectronic Systems, and School of Physics, Beijing Institute of Technology, Beijing 100081, China}

\author{Yugui Yao}
\affiliation{Key Lab of advanced optoelectronic quantum architecture and measurement (MOE), Beijing Key Lab of Nanophotonics $\&$ Ultrafine Optoelectronic Systems, and School of Physics, Beijing Institute of Technology, Beijing 100081, China}

\begin{abstract}
Recently, the real topology has been attracting widespread interest in two dimensions  (2D).
Here, based on  first-principles calculations and theoretical analysis, we reveal the  monolayer  Cr$_2$Se$_2$O (ML-CrSeO) as the first material example of a 2D antiferromagnetic (AFM) real Chern insulator (RCI) with  topologically protected corner states.
Unlike previous RCIs,  we find that the real topology of the  ML-CrSeO is  rooted  in one certain  mirror subsystem of the two  spin channels, and can not be directly obtained from  all the valence bands in each spin channel  as  commonly believed.
In particular, due to  antiferromagnetism, the corner modes in ML-CrSeO exhibit strong  corner-contrasted spin polarization, leading to spin-corner coupling (SCC). This SCC enables a direct connection  between spin space and real space.
Consequently, large and switchable net magnetization can be induced in the ML-CrSeO nanodisk by  electrostatic means, such as potential step and in-plane electric field, and the corresponding  magnetoelectric responses behave like  a sign function, distinguished from that of the conventional multiferroic materials.
Our work considerably broadens the candidate range of RCI  materials, and  opens up a new direction for  topo-spintronics and 2D AFM materials research.
\end{abstract}

\maketitle

Materials with nontrivial real band topology~\cite{PhysRevLett.118.056401,PhysRevLett.121.106403,wu2019non,luo2022fragile} have become a focus of current physics research.
As a definition, the real topological phases feature real band eigenstates~\cite{PhysRevLett.118.056401,PhysRevLett.121.106403}, which can be guaranteed by certain symmetries, such as the presence of the spacetime inversion symmetry (${\cal{PT}}$) and the absence of spin-orbit coupling (SOC) effect.
Furthermore, for 2D systems both with and without SOC, the combined operator ${C_{2z}\cal{T}}$ (with $z$ being the direction normal to the 2D plane) can also protect the real band  topology~\cite{PhysRevB.99.235125,PhysRevB.105.085123}.

Various real topological phases are proposed in nonmagnetic systems without SOC, including 2D and 3D RCIs, Z$_2$ nodal lines,  Z$_2$ nodal surfaces, real  Weyl and Dirac  points~\cite{PhysRevB.97.115125,PhysRevLett.121.106403,PhysRevLett.123.256402,zhao2020equivariant,PhysRevLett.125.126403,PhysRevB.104.085205,PhysRevLett.128.026405,pan2022two,PhysRevB.105.085123,pan2023real,xue2023stiefel,xiang2023demonstration,10.1063/5.0159948,doi:10.1021/acs.nanolett.3c01723}, and some of them have been experimentally realized~\cite{pan2023real,xue2023stiefel,xiang2023demonstration}.
Recently, Zhang \emph{et. al.}~\cite{doi:10.1021/acs.nanolett.3c01723} demonstrated that the 2D RCI could also appear in ferromagnetic systems regardless of the SOC strength.
However, the 2D RCI with AFM order has never been studied. Particularly, due to the magnetic but  spin neural property, one can expect that the AFM RCI may exhibit  signatures distinct from the conventional and ferromagnetic RCIs, especially for the magnetoelectric responses~\cite{ma2021multifunctional,zhang2023predictable}.

Generally, the real topology of a 2D insulator is characterized by a $\mathbb{Z}_2$ real Chern number  (also known as the second Stiefel-Whitney number) $\nu_R$~\cite{PhysRevLett.118.056401,PhysRevLett.121.106403}, which  can be evaluated from the eigenvalues of two-fold rotation $C_{2z}$ or spatial inversion  ${\cal{P}}$  of the valence bands at the four $C_{2z}$ (${\cal{P}}$) invariant momentum points $\Gamma_i$ ($i=1,2,3,4$)~\cite{PhysRevLett.121.106403,ahn2019stiefel}
\begin{equation} \label{RCI_num}
(-1)^{\nu_{\mathrm{R}}}=\prod_{i=1}^4(-1)^{\left\lfloor\left(n_{i,-}\right) / 2\right\rfloor},
\end{equation}
where $\lfloor\cdots\rfloor$ is the floor function, and $n_{i,-}$ is the number of valence states at four momentum points $\Gamma_i$ which have negative eigenvalues.
This formula (\ref{RCI_num}) successfully predicts many RCI and topological Z$_2$  nodal line materials~\cite{PhysRevLett.128.026405,PhysRevB.104.085205,PhysRevLett.123.256402,PhysRevB.105.085123,10.1063/5.0159948,pan2022two,PhysRevLett.125.126403,pan2023real,xue2023stiefel,xiang2023demonstration,10.1063/5.0159948,doi:10.1021/acs.nanolett.3c01723}.

In this work, we show that there is a caveat for calculating $\nu_R$ of the  system that can be divided into two or multiple \emph{independent} subsystems,  due to the floor function.
For example, assuming  the numbers of the valence states with negative $C_{2z}$ eigenvalue in two independent subsystems are respectively $n^{(1)}_{-}=\{6,6,5,4\}$, and $n^{(2)}_{-}=\{4,4,5,6\}$ at   the four $C_{2z}$-invariant points.
For each subsystem, one can obtain a real Chern number by Eq. (\ref{RCI_num}), which gives $\nu_R^{(1)}=0$ and $\nu_R^{(2)}=1$.
Therefore, the whole system is  topologically nontrivial.
However, this real topology is missed if one directly applies  Eq. (\ref{RCI_num}) to the whole system, as  $n_{-}=n^{(1)}_{-}+n^{(2)}_{-}=\{10,10,10,10\}$, which gives a trivial real Chern number  $\nu_R=0$.
This simple case tells us that for the materials hosting  multiple independent subsystems, since Eq. (\ref{RCI_num}) does not naturally guarantee $\nu_R=\sum_{i}\nu_R^{(i)} \ \text{mod} \  2$ (with $\nu_R^{(i)}$ denoting the real Chern number of the $i$-th  subsystem), the real topology of the whole system may be hidden in one certain subsystem, and needs to be carefully examined.

Via first-principles calculations and theoretical analysis, we demonstrate the aforementioned hidden real topology in   ML-CrSeO,   which  has a  horizontal mirror $M_z$, and then two independent  mirror subsystems.
Particularly, the ground state of the  ML-CrSeO  exhibits out-of-plane intralayer AFM  order. This means that  the  ML-CrSeO is also the  first material example of a 2D AFM RCI.
Compared with the previously reported nonmagnetic and ferromagnetic RCIs~\cite{PhysRevLett.123.256402,PhysRevB.105.085123,PhysRevB.104.085205,10.1063/5.0159948,pan2022two,doi:10.1021/acs.nanolett.3c01723}, the topological corner states in   ML-CrSeO are  spin-polarized, however,  the direction of the spin polarization is corner-dependent, leading to a unique SCC effect in 2D AFM RCIs.
The SCC can be directly  observed by spin-resolved scanning tunneling spectroscopy (STS)~\cite{RevModPhys.81.1495}.
Furthermore, it  can  fundamentally affect the  magnetoelectric responses of the systems.
We find that by applying  electrostatic fields, the  net magnetization of the  ML-CrSeO can be efficiently changed from zero to a finite constant.
Remarkably, the sign of the   finite constant is solely determined by the direction  of the electrostatic fields, and is  switched when changing  the  direction of the    fields.
Hence, the magnetoelectric responses in  ML-CrSeO behave as   a sign function, distinguished from those in the conventional multiferroic materials~\cite{eerenstein2006multiferroic} and the previously reported  RCIs.
This sign-function response also indicates an electric switching of the net magnetization in ML-CrSeO, which is important for device applications~\cite{matsukura2015control,wadley2016electrical,song2017recent}.

\begin{figure}[t]
\includegraphics[width=8.5cm]{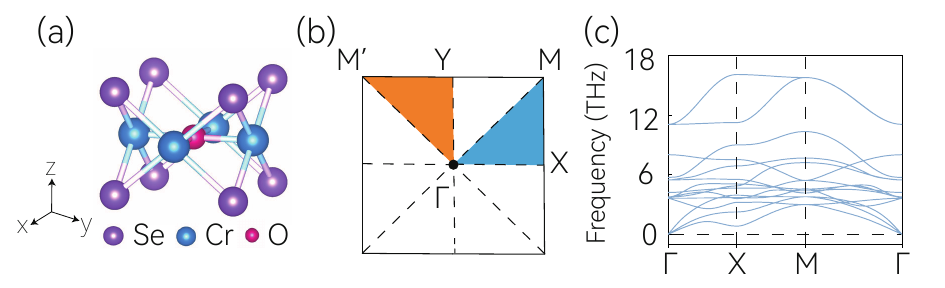}
\caption{ (a) Side view of the crystalline structure of ML-CrSeO. (b) The first Brillouin zone (BZ) of ML-CrSeO. The orange and blue regions are the two irreducible parts of the first BZ. (c) Calculated phonon spectrum of  ML-CrSeO.
\label{fig1}}
\end{figure}

{\emph{\textcolor{blue}{Structure and magnetic ordering.--}}}
The  ML-CrSeO is  a kind of oxyselenides.
The crystal structure of the   ML-CrSeO is shown in Fig.~\ref{fig1}(a), in which three atomic sublayers with two Cr and one O atoms as the middle layer and two Se atoms as the upper and lower layers are obvious.
The crystal lattice has space group (SG) P4/mmm  (No. 123), which preserves $C_{4z}$, $C_{2x}$ and $M_z$ symmetries. The optimized lattice constant is calculated as $a = b = 4.02$  \AA.
We also find that the ML-CrSeO is both  dynamically and thermodynamically stable, as shown in Fig.~\ref{fig1}(c) and the supplemental material (SM)~\cite{SM}.

\begin{figure}[t]
\includegraphics[width=8.5cm]{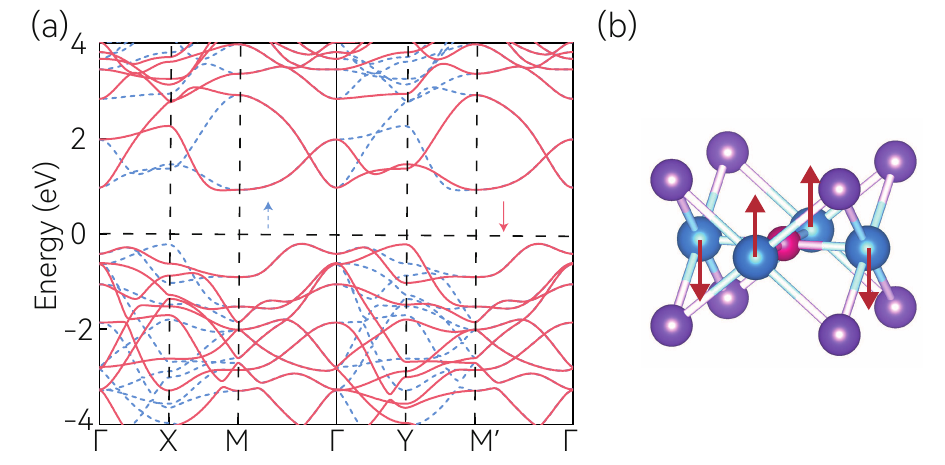}
\caption{(a) Spin-resolved band structures for ML-CrSeO without SOC. (b) The ground state of the ML-CrSeO, which is a intralayer out-of-plane (C-type) AFM.
\label{fig2}}
\end{figure}

Since the ML-CrSeO has transition metal elements, one can  expect that it may exhibit magnetic ordering in its ground state.
Our calculations show each Cr atom indeed has a finite magnetic moment, which is obtained as $\sim3.1~\mu{_B}$.
By comparing the energies of several typical types of magnetic ordering~\cite{SM}, we find that the ground state of ML-CrSeO is a C-type AFM with intralayer out-of-plane magnetization, i.e., the two Cr atoms in the unit cell have opposite spin polarization, as shown in Fig.~\ref{fig2}(b).
With magnetic ordering, the ML-CrSeO  belongs to magnetic SG No. 123.342, which keeps $C_{4z}{\cal{T}}$, $C_{2z}$ and $M_z$ symmetries, but breaks both $C_{4z}$ and ${\cal{T}}$~\cite{PhysRevB.105.085117,yu2022encyclopedia,PhysRevB.105.104426}.
The combined operator $C_{4z}{\cal{T}}$ guarantees the spin neutrality of the monolayer.

{\emph{\textcolor{blue}{Electronic band and hidden real topology.--}}}
We then investigate the  electronic band structure of  the ML-CrSeO with the AFM ground state.
Since the SOC effect is rather small in the ML-CrSeO~\cite{SM}, we focus on the band structure without SOC.

When SOC is neglected, the two spin channels of the ML-CrSeO are decoupled but  not independent, as they  are connected by $C_{4z}{\cal{T}}$.
Each spin channel can be regarded as an effective spinless system persevering ${\cal{T}}$.
Obviously, each spin channel does not have $C_{4z}$ symmetry and then the crystal symmetry of SG 123.
By detailed analysis, we find that it is the crystal symmetry of SG 47 that is respected by each spin channel.
The generators of SG 47 are  $C_{2x}$, $C_{2z}$ and $M_z$.
Hence, the Hamiltonian of the ML-CrSeO (without SOC) can be written as
\begin{equation}
{\cal{H}}={\cal{H}}_{\uparrow} \oplus {\cal{H}}_{\downarrow}={\cal{H}}_{0}\oplus C_{4z}{\cal{H}}_0C_{4z}^{-1},
\end{equation}
where ${\cal{H}}_{0}$ denotes a spinless and nonmagnetic Hamiltonian respecting the symmetry of SG 47.
Since  ${\cal{H}}_{0}$ has $C_{2x}$ and ${\cal{T}}$, one knows that the two spin channels should be degenerate along $\Gamma$-M path.

\begin{table}[b]
\caption{\label{Tab1} $C_{2 z}$ eigenvalues of  all occupied bands at $\Gamma$, X, Y, and M points of the ML-CrSeO. }
\begin{ruledtabular}
\begin{tabular}{cccccccccccc}
  \multicolumn{5}{c}{Spin-up}   &  \multicolumn{5}{c}{Spin-down} \\
       \hline
    & $\Gamma$ & X & Y & M & && $\Gamma$ & X & Y & M  \\
    $n_+$ & 10 & 10 & 11 & 11 && $n_+$ & 10 & 11 & 10 & 11  \\
    $n_-$ & 11 & 11 & 10 & 10 && $n_-$ & 11 & 10 & 11 & 10 \\
       &  & $\nu_{\mathrm{R}}$ = 0  && & &  &  & $\nu_{\mathrm{R}}$ = 0  &  &   \\
\end{tabular}
\end{ruledtabular}
\end{table}

%
%

\begin{table*}
\caption{\label{Tab2} Mirror-resolved $C_{2 z}$ eigenvalues of all occupied bands at $\Gamma$, X, Y, and M points of the ML-CrSeO. Both spin channels are considered. }
\begin{ruledtabular}
\begin{tabular}{ccccccccccccccccccccc}

\multicolumn{10}{c}{Spin-up} &  &  & \multicolumn{9}{c}{Spin-down}\tabularnewline
\hline
 & \multicolumn{4}{c}{$M_{z}=1$} &  & \multicolumn{4}{c}{$M_{z}=-1$} &  &  & \multicolumn{4}{c}{$M_{z}=1$} &  & \multicolumn{4}{c}{$M_{z}=-1$}\tabularnewline
 & $\Gamma$ & X & Y & M &  & $\Gamma$ & X & Y & M &  &  & $\Gamma$ & X & Y & M &  & $\Gamma$ & X & Y & M\tabularnewline
$n_{+}$ & 5 & 6 & 8 & 7 &  & 5 & 4 & 3 & 4 &  & $n_{+}$ & 5 & 8 & 6 & 7 &  & 5 & 3 & 4 & 4\tabularnewline
$n_{-}$ & 8 & 7 & 5 & 6 &  & 3 & 4 & 5 & 4 &  & $n_{-}$ & 8 & 5 & 7 & 6 &  & 3 & 5 & 4 & 4\tabularnewline
 & \multicolumn{4}{c}{$\nu_{R}=0$} &  & \multicolumn{4}{c}{$\nu_{R}=1$} &  &  & \multicolumn{4}{c}{$\nu_{R}=0$} &  & \multicolumn{4}{c}{$\nu_{R}=1$}\tabularnewline
\end{tabular}
    \end{ruledtabular}
\end{table*}

The spin-resolved  band structures for the ML-CrSeO  are shown in Fig.~\ref{fig2}(a), in which the spin-up and spin-down bands are degenerate on the high-symmetry line $\Gamma$-M (M'), consistent with the above symmetry analysis.
Besides, the  spin-up and spin-down bands are split at other momenta in BZ.
Recently, the systems with spin-splitting  band structures are also termed  as  altermagnetic materials~\cite{PhysRevX.12.040501, feng2022anomalous,vsmejkal2022anomalous,PhysRevX.12.031042,PhysRevLett.130.216701,PhysRevLett.131.076003}.
For both spin channels, large gaps of 1.12 eV appear.
Then, we can define a real Chern number $\nu_R^\sigma$ (with $\sigma=\uparrow,\downarrow$) for each spin subspace, as it preserves both $C_{2z}$ and ${\cal{T}}$ symmetries.
Since $C_{4z}$ commutes with $C_{2z}$, one always  has $\nu_R^\uparrow=\nu_R^\downarrow$.
The results of the $C_{2z}$ eigenvalues of valence bands at the four $C_{2z}$-invariant momentum points are listed in  Table \ref{Tab1}.
Obviously, according to Eq. (\ref{RCI_num}), both spin channels are topologically trivial with  $\nu_R^\uparrow=\nu_R^\downarrow=0$.
However, as aforementioned, this result may not be correct  if the spin channels can be further  divided into independent  subsystems.

\begin{figure}
\includegraphics[width=8.7cm]{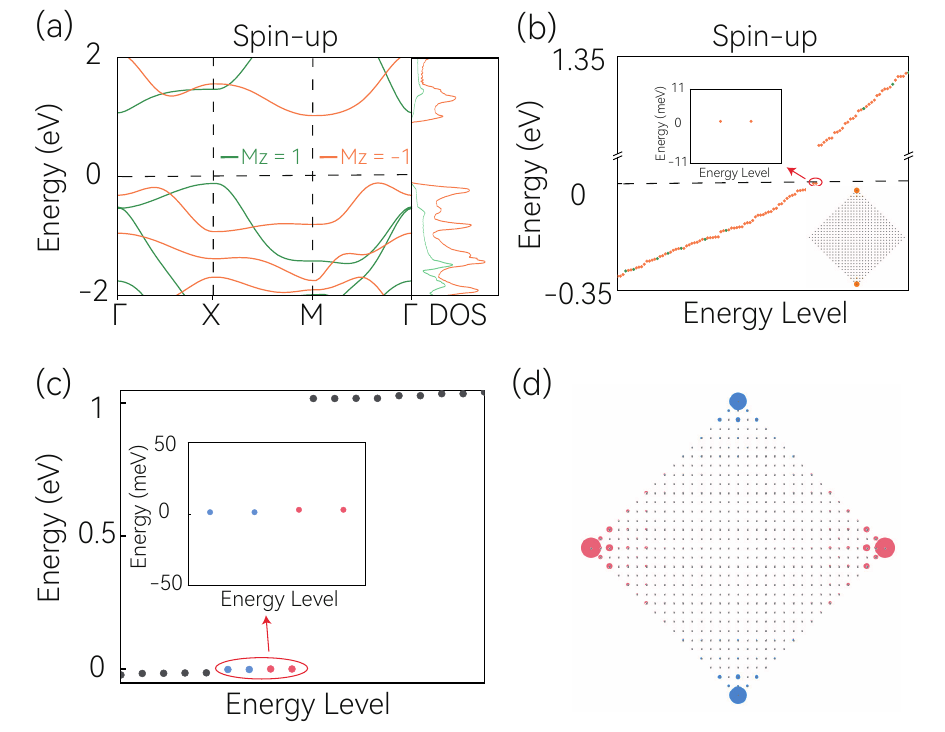}
\caption{(a) Mirror-resolved band structure and density of states (DOS) of the spin-up channel of the ML-CrSeO. The spin-down channel has a similar band structure, as the two spin channels are connected by $C_{4z}{\cal{T}}$ symmetry.  (b) Mirror-resolved  energy spectrum for the spin-up channel  on a quadrangle-shaped disk. Orange and green curves and dots in (a) and (b) denote the states with  $M_z=-1$ and $M_z=1$, respectively.
We also plot the spectrum and the spatial charge distribution of the corner states as insets in (b).
(c) The energy spectrum for  the whole  ML-CrSeO on a quadrangle-shaped disk, where four corner states (the colored  dots) as a group can be observed. (d) Spatial distribution of the corner states. The blue and red dots (circles) in (c) [(d)] correspond to the spin-up and spin-down corner states, respectively.
\label{fig3}}
\end{figure}

The ML-CrSeO is exactly the case, as it has a horizontal mirror $M_z$ and then can be  divided  into two independent subsystems based on the  eigenvalue of $M_z$, i.e.,
\begin{equation}
{\cal{H}}_{\uparrow(\downarrow)}={\cal{H}}_{\uparrow(\downarrow)}^{+} \oplus {\cal{H}}_{\uparrow(\downarrow)}^{-},
\end{equation}
where  ${\cal{H}}_{\uparrow(\downarrow)}^{\pm}$  denotes the  subsystem with $M_z=\pm1$ of spin-up (spin-down) channel.
All the mirror subsystems  also have both $C_{2z}$ and ${\cal{T}}$.
By analysing the mirror eigenvalue of the bands in each spin channel, we  obtain a mirror-resolved band structure for the ML-CrSeO [see Fig.~\ref{fig3}(a)].
For each mirror subsystem, we count the bands with negative $C_{2z}$ eigenvalue, and the result is listed in Table \ref{Tab2}.
Using  Eq. (\ref{RCI_num}), we find that  for both spin-up and spin-down channels, the  mirror subsystem with $M_z=-1$ has a nontrivial $\nu_R^{-}=1$, while the $M_z=1$ subsystem is trivial with $\nu_R^{+}=0$.
Consequently, the \emph{exact} real Chern number for each spin channel is $\nu_R^\uparrow=\nu_R^\downarrow=\nu_R^{+}+\nu_R^{-}=1$.
This confirms that the  ML-CrSeO is a material candidate for AFM RCI with  hidden real topology.

{\emph{\textcolor{blue}{Corner states and spin-corner coupling.--}}}
According to the bulk-boundary correspondence, the nontrivial real Chern number $\nu_R^{-}=1$ will lead to protected corner  states on a pair of $C_{2z}$-related  intersections in each  spin channel, and the corner states should have a mirror eigenvalue of $M_z=-1$.
Moreover, since the spin-up and spin-down channels are connected by $C_{4z}{\cal{T}}$, there genetically exhibit four corner states as a group in the bulk band gap: two from spin-up channels and two from spin-down channels.

To demonstrate this feature, a wannier tight-binding model for a quadrangle nanodisk composed of 64 unit cells was constructed based on the ML-CrSeO with the preservation of $C_{4z}{\cal{T}}$  symmetry, as shown in Fig.~\ref{fig3}(d).
The obtained discrete energy spectrum of the quadrangle nanodisk is plotted in Fig.~\ref{fig3}(c), in which four degenerate states in the  bulk band gap  are observed.
By analysing the wave function of the four states, we confirm that they are indeed corner states, as they are localized at the four corners of the disk [see Fig. \ref{fig3}(d)].
Moreover, all these corner states are from the $M_z=-1$ subsystems [see Fig.~\ref{fig3}(b)] and  are  fully spin polarized, as the SOC effect is  neglected during the calculations.
We would like to point out that both the existence and the spin polarization of  the corner states are robust against the SOC effect \cite{SM}.

Remarkably, the  corner states at $x$-axis (labelled as X corner) are spin-down, whereas those at $y$-axis (labelled as Y corner) are spin-up, as shown in Fig.~\ref{fig3}(d).
This corner-contrasted spin polarization directly couples the real space and spin space,  leading to a novel SCC effect.
Since the corner states are well separated in  real space (for example, the distance between every two corners in Fig.~\ref{fig3}(d) is larger than  $3.2$ nm),
this  provides great convenience for experimentally controlling the spins at each corner independently.
Moreover, as electronic states at the Fermi level  play a key role in many material properties, the independent control of the spin in each corner is  expected to exhibit distinctive  signatures.

\begin{figure}
\includegraphics[width=8.5cm]{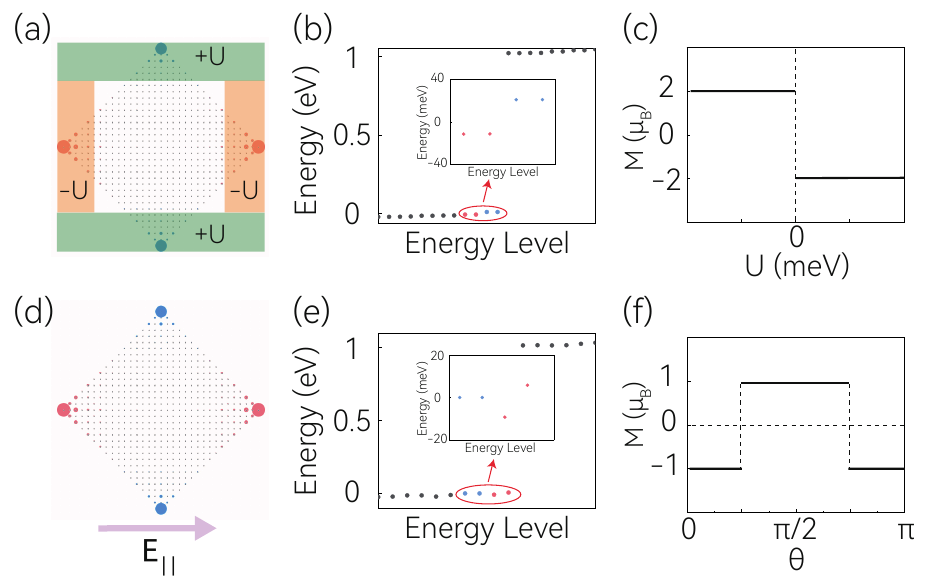}
\caption{SCC-induced electric control of the magnetization. (a) Schematic figure showing potential-step control of the magnetization of the ML-CrSeO disk. (b) The energy spectrum for the ML-CrSeO disk under a potential step $U=0.01$ eV. (c) Net magnetization as a function of $U$. (d) With corner-dependent spin polarization, an in-plane electric field can lead to net magnetization in the ML-CrSeO disk. (e) The energy spectrum for the  ML-CrSeO disk  under an electric field of $E_{x}=0.00025$ eV/\AA, corresponding to $\theta=0$. (f) Net magnetization as a function of  $\theta$. The blue and red circles (dots) are spin-up and spin-down corner states.
\label{fig4}}
\end{figure}

{\emph{\textcolor{blue}{Novel magnetoelectric responses.--}}}
We then investigate  the magnetoelectric responses in the ML-CrSeO.
A  unique feature of AFM materials is that it is magnetic but spin-neutral~\cite{RevModPhys.90.015005,yan2020electric}. This provides an ideal platform for generating net magnetization and spin  current by nonmagnetic methods.
As studied in previous works~\cite{PhysRevLett.118.106402,ma2021multifunctional,shao2021spin,PhysRevLett.126.127701,zhang2023predictable,PhysRevLett.130.216702}, these novel magnetoelectric responses generally require certain bulk band structures like spin-layer coupling~\cite{zhang2023predictable} and C-paired spin-valley locking~\cite{ma2021multifunctional}, which, however, are absent in the ML-CrSeO.

In contrast, we investigate the magnetoelectric responses of a ML-CrSeO nanodisk with a novel SCC effect.
While the corner states in ML-CrSeO are fully spin polarized, the nanodisk does not have net magnetization as guaranteed by the $C_{4z}{\cal{T}}$ symmetry.
Hence, by breaking  the $C_{4z}{\cal{T}}$, the ML-CrSeO nanodisk generally has  finite  net magnetization.
There are many nonmagnetic ways to realize this symmetry breaking, such as uniaxial strain,  applying nonmagnetic substrate without  $C_{4z}$ symmetry and interface engineering.
Among them, the electrostatic means may be the most convenient in experiments and applications~\cite{PhysRevLett.124.037701}.
Here, we propose  two  electrostatic means  to achieve net  magnetization in the ML-CrSeO nanodisk.

As the four corners of the nanodisk are well separated, we can apply a potential step to  make each corner to have different potential energy.
Specifically, we assume that the potential step satisfies $V=-U$ for the X corners and  $V=+U$ for the Y corners, as illustrated in Fig.~\ref{fig4}(a).
The calculated spectrum of the nanodisk with $U = 0.01$ eV is shown in Fig.~\ref{fig4}(b).
As expected, the degenerate of the X and Y corners is broken.
The Y (X) corners increase (decrease) in energy, but the bulk states are less affected.
Hence, with half doping of the corner states, only the X corners are occupied, and then the net magnetization of the system is $M=-2\mu_B$.
Moreover, the net  magnetization can be switched by reversing the value of $U$.
Consequently, the net  magnetization takes the form of a sign function by varying the electrostatic potential, as shown in Fig.~\ref{fig4}(c).

We can also use in-plane electric field ${{\bm E}}_{\parallel}=\{E \cos\theta, E\sin \theta\}$ to produce  a corner-dependent electrostatic potential [see Fig.~\ref{fig4}(d)].
When the electric field is along the $x$-direction ($\theta=0$), the four corner states are  split into three  groups in energy: the -X (+X) corner has the lowest (highest) energy, while  the two Y corners are in the middle [see Fig.~\ref{fig4}(e)].
Then, with $1/4$ doping of the corner states, only the -X corner is occupied, and then the net magnetization of the system is $M=-1\mu_B$.
Similarly, if $\theta=\pi/2$, one would have $M=1\mu_B$.
Our detailed calculations show  that the net magnetization  has a sign-function dependence on the rotation  angle of the electric field, as shown in Fig.~\ref{fig4}(f).
These sign functions of the magnetoelectric responses are completely different from the conventional magnetoelectric responses, suggesting the potential of the AFM RCIs for novel device operations.

{\emph{\textcolor{blue}{Discussion.--}}}
In this work, we have unveiled  a hidden real topology in a 2D AFM material  with mirror symmetry. The underlying physics is to study the real topology in each independent subsystem, rather than the whole system.
We then  predict the ML-CrSeO as the first material example of the AFM RCI with hidden real topology.
We also show that the 2D AFM RCI exhibits the SCC effect, which  enables a  direct control of the net magnetization of the system by both local and global electric methods.

Besides the ML-CrSeO, we find six other 2D AFM materials, including Cr$_2$Br$_2$, Mn$_2$Br$_2$, V$_2$Br$_2$, Mn$_2$F$_2$, Cr$_2$Cl$_2$, and V$_2$Se$_2$O as  the AFM RCIs with hidden real topology~\cite{SM}.
Interestingly, the Cr$_2$Br$_2$, Cr$_2$Cl$_2$, and V$_2$Se$_2$O are also valleytronics  materials~\cite{schaibley2016valleytronics,li2019facile,chu20212d,liu2023strain} with two valleys  located at X and Y points in the BZ.
Last but not least, it is necessary to re-examine the previous 2D materials with a horizontal mirror.
With the hidden real topology proposed here, one can expect that many candidates  previously diagnosed as trivial may have a nontrivial real Chern number.




\bibliography{refSI}

\end{document}